%% file: main_ieeevisshort.tex
\documentclass{vgtc}                          %

\ifpdf%
  \pdfoutput=1\relax                   %
  \usepackage{graphicx}                %
  \DeclareGraphicsExtensions{.pdf,.png,.jpg,.jpeg} %
\else%
  \usepackage{graphicx}                %
  \DeclareGraphicsExtensions{.eps}     %
\fi%

\graphicspath{{figures/}{pictures/}{images/}{./}} %

\usepackage{microtype}                 %
\PassOptionsToPackage{warn}{textcomp}  %
\usepackage{textcomp}                  %
\usepackage{mathptmx}                  %
\usepackage{times}                     %
\usepackage{cite}                      %
\usepackage{tabu}                      %
\usepackage{booktabs}                  %
\usepackage{tabularx}

\usepackage{amsmath}
\usepackage{tabularx}
\usepackage[table]{xcolor}
\usepackage{multirow}
\usepackage{threeparttable}

\newcommand{\hint}[1]{\textcolor{red}{#1}}
\newcommand{\revision}[1]{\leavevmode{\color{blue}{#1}}}
\newcommand{\removed}[1]{\leavevmode{\color{red}{\st{#1}}}}

\def \cleanversion{} %
\ifx\cleanversion\undefined
\else
 \renewcommand{\hint}[1]{}
 \renewcommand{\removed}[1]{} 
 \renewcommand{\revision}[1]{#1}
\fi

\newcommand{\datasetname}{SimVecVis}

\usepackage{mdframed}

\usepackage{enumitem}

\usepackage{fvextra} %

\DefineVerbatimEnvironment{MyVerb}{Verbatim}{
  breaklines=true,
  breakanywhere=true
}

\usepackage{caption}
\setlist[itemize]{leftmargin=8pt, itemsep=0pt, topsep=2pt}
\setlist{noitemsep,parsep=0pt,partopsep=0pt}

\vgtccategory{Research}

\vgtcinsertpkg

\input{meta/commands}

\title{SimVecVis: A Dataset for Enhancing MLLMs in Visualization Understanding}

\author{Can Liu\thanks{Equal contribution. E-mail: can.liu@ntu.edu.sg}\\ %
        \scriptsize Nanyang Technological University %
\and Chunlin Da\thanks{Equal contribution. E-mail: dachunlin@bytedance.com}\\ %
     \scriptsize ByteDance Inc. %
\and Xiaoxiao Long\thanks{E-mail: xiaoxiao.long@nju.edu.cn}\\ %
     \parbox{1.4in}{\scriptsize \centering Nanjing University}
\and Yuxiao Yang\thanks{E-mail: yuxiao@fondant.design}\\
    \scriptsize Tsinghua University
\and Yu Zhang\thanks{E-mail: yu.zhang@cs.ox.ac.uk}\\
    \scriptsize University of Oxford
\and Yong Wang \thanks{Corresponding author. E-mail: yong-wang@ntu.edu.sg}\\
    \scriptsize Nanyang Technological University
}

\abstract{
    Current multimodal large language models (MLLMs), while effective in natural image understanding, struggle with visualization understanding due to their inability to decode the data-to-visual mapping and extract structured information.
    To address these challenges, we propose SimVec,
    a novel
    simplified %
    vector format that encodes chart elements such as mark type, position, and size. The effectiveness of SimVec is demonstrated by using MLLMs to reconstruct chart information from SimVec formats. 
    Then, we build a new visualization dataset, SimVecVis, to enhance the performance of MLLMs in visualization understanding, which consists of three key dimensions: bitmap images of charts, their SimVec representations, and corresponding data-centric question-answering (QA) pairs with explanatory chain-of-thought (CoT) descriptions.
    We fine-tune state-of-the-art MLLMs (e.g., MiniCPM and Qwen-VL), using SimVecVis with different dataset dimensions.
    The experimental results show that it leads to substantial performance improvements of MLLMs with good spatial perception capabilities (e.g., MiniCPM)
    in data-centric QA tasks.
    Our dataset and source code are available at: \textcolor{blue}{\url{https://github.com/VIDA-Lab/SimVecVis}}.
    
} %

\CCScatlist{
  \CCScatTwelve{Human-centered computing}{Visualization}{}{}
}


\begin{document}

\maketitle
\input{sections/1_introduction}

\input{sections/2_related_work}
\input{sections/3_dataset}

\input{sections/4_model}
\input{sections/5_experiments}

\input{sections/6_conclusion}

\acknowledgments{
This project is supported by the Ministry of Education, Singapore, under its Academic Research Fund Tier 2 (Proposal ID: T2EP20222-0049).
Any opinions, findings and conclusions, or recommendations expressed in this material are those of the author(s) and do not reflect the views of the Ministry of Education, Singapore.
}

\clearpage

\bibliographystyle{abbrv-doi}

\bibliography{main}
\clearpage
\appendix

\end{document}

%% file: meta/commands.tex
\usepackage{soul}

%% file: sections/1_introduction.tex
\section{Introduction}

\removed{Visualizations have emerged as critical tools for comprehending and conveying complex information.
Visualization empowers decision-makers, analysts, and the general public to interpret data more efficiently, leading to informed decisions.
The relevance of visualizations is evident across a spectrum of fields, including business, scientific research, education, and public policy.}
\revision{As charts have become a dominant medium for conveying data in scientific and practical contexts, manual interpretation at scale has become infeasible, calling for automated methods that can reliably understand visualizations.
However, current MLLMs, originally designed for natural images, fall short in this domain due to a fundamental distinction: natural images depict real-world objects based on their visual appearance, while visualizations convey
data through encoding rules that map data to visual attributes of elements such as marks, axes, and legends.
Existing MLLMs are often not equipped to interpret such encoding rules, which are rarely present in natural image training data.}
\removed{However, due to the significant differences between visualization images and natural images, existing methods for chart understanding are still inadequate.
The reason is that natural images capture concrete objects in real-world scenarios, while visualizations operate through mapping rules to represent data, providing a means to convey more nuanced and complex information.
This distinction requires that large models not only excel in accurate content recognition but also preserve the integrity of visual encoding to ensure chart accuracy and readability.}

\removed{Additionally, the key insights conveyed in charts are often more refined and focused than those obtained through underlying data.
Moreover, supporting chart analysis can also advance multimodal models in their ability to understand complex data.
Chart interpretation represents a fine-grained challenge within multimodal tasks.
In light of these limitations, there is an urgent need for models capable of interpreting visualizations with greater accuracy.}
\revision{Visualization understanding
goes beyond the mere recognition of visible content.
It involves reasoning about how visual channels represent data values and inferring the underlying data in visualizations from different sources, including scanned documents and historical print media.}\removed{Existing large models often fail to capture the mapping rules and detailed data necessary for understanding chart representations, thereby limiting their applicability in chart comprehension and analysis.}
\revision{
To address the challenge of visualization understanding,
we first propose SimVec, a novel \textbf{Sim}plified \textbf{Vec}tor) format to capture visual mark attributes (e.g., mark type, position, size, color), which provides a machine-readable abstraction of visualizations.
Built upon it, we further construct a new visualization dataset, \datasetname{}, for fine-tuning MLLMs and improving their performance of visualization understanding.
\datasetname{} contains 2,999 visualizations like bar charts, line charts, and area charts, and each visualization consists of three key dimensions: visualization bitmap image, SimVec representation, and data-centric QA pairs with CoT descriptions.
The visualization bitmap image is commonly used for visualization understanding tasks and encodes the comprehensive information of a chart.
The corresponding SimVec representation provides a more machine-friendly encoding of visualization information. The data-centric QA pairs with CoT descriptions cover chart QA tasks, like identifying the value of the tallest bar in a bar chart, and include detailed CoT reasoning descriptions, which can guide MLLMs to learn the correct way to do \textbf{visualization understand} as well as reasoning. 
We conduct extensive experiments of fine-tuning MLLMs with \datasetname{} with different dataset dimensions, and the results demonstrate its effectiveness in significantly enhancing the visualization understanding performance of MLLMs with good spatial perception capabilities like MiniCPM~\cite{hu2024minicpm}.
The expressiveness of our SimVec representations is also proved via visualization information construction experiments.
}

The contributions of this work are summarized as follows:

\begin{itemize}
    \item We propose a novel \revision{chart} format, SimVec, for a compact and structured representation of
    charts.

    \item \revision{We construct a visualization dataset with 2,999 visualizations, \datasetname{}, 
    aiming to explicitly enhance MLLMs' visualization understanding capabilities. 
    }

    \item We show the expressiveness of SimVec via MLLM-based chart information reconstruction
    and demonstrate that fine-tuning MLLMs with \datasetname{} can improve their performance in visualization understanding.

\end{itemize}

%% file: sections/2_related_work.tex
\section{Related Work}

\subsection{Artificial Intelligence for Visualization Understanding}

Artificial intelligence has increasingly contributed to improving the understanding of visualizations.
Early systems~\cite{cox2001multi, mackinlay2007showme, sun2010articulate} were primarily rule-based, leveraging handcrafted grammars to parse chart structures and natural language.
For example, Show Me~\cite{mackinlay2007showme}, Voyager~\cite{wongsuphasawat2016voyager}, Iris~\cite{fast2018Iris}, and FlowSense~\cite{yu2020flowsense} applied structured rules to support visualization querying.
With the rise of deep learning, approaches~\cite{moritz2018formalizing_draco, luo2018deepeye, liu2021advisor} model the relationship between visual encodings and semantics, enabling more accurate recommendations.
A growing body of work focuses on understanding of existing visualizations through natural language. Key tasks include chart question answering (ChartQA)~\cite{kushal2018dvqa, kim2020answering, wei2024mchartqa, samira2018figureqa}, chart captioning~\cite{chen2020figurecap, liu2020autocaption}, and automatic natural language annotation~\cite{Lai2020Annotation}.
These tasks aim to convert chart content into human-readable forms to facilitate interpretation.

Recently, MLLMs have been applied to visualization understanding.
For example, mChartQA~\cite{wei2024mchartqa} transforms charts into data tables to enable precise chart reasoning, while systems such as TinyChart~\cite{zhang2024tinychart} and ChartX~\cite{xia2024chartx} directly process chart images through vision-language models.
Despite promising results, visualizations often involve spatially structured elements, such as axes, tick labels, and marks, which are not well represented in general pretraining.
Moreover, current MLLMs for visualization tend to produce direct outputs without explicit reasoning, unlike human users, who typically engage in step-by-step inference when interpreting visualizations.
To address these gaps, we propose SimVecVis, which introduces a compact vectorized representation of charts and incorporates chain-of-thought (CoT) reasoning into chart question answering. We further fine-tune our framework on models such as MiniCPM~\cite{hu2024minicpm}, which demonstrate good spatial awareness and support high-resolution visual input.

\subsection{Visualization Datasets}

Visualization researchers have constructed a series of visualization datasets~\cite{liu2024dataset} in the past few years.
For example, VisImages~\cite{deng2022visimages} compiles visualizations extracted from diverse media.
D3 search~\cite{d3search} crawled specific websites to collect visualizations created with D3~\cite{bostock2011d3}, while VizML~\cite{vizml} retrieves data and visual encoding specifications from an online gallery.
VizNet~\cite{viznet} offers a large-scale corpus of 31 million data points gathered from open data repositories and online galleries.
The reverse engineering visualization dataset~\cite{poco2017reverse} aggregates images from Vega Charts, news sites, and academic papers.
Fu et al.~\cite{fu2019visualization} use dimension reduction technique to derive vector representations from infographic images.
MASSVIS~\cite{memorable} automatically collects visualizations across fields from online websites.
OldVisOnline and ZuantuSet gather historical visualizations before the computer era~\cite{zhang2024oldvisonline,Mei2025ZuantuSet}.
However, existing datasets overlook reasoning processes such as CoT and lack compact vector representations for reconstruction, which are the focus of \datasetname{}.

%% file: sections/3_dataset.tex
\begin{figure}
    \centering
    \includegraphics[width=1.0\linewidth]{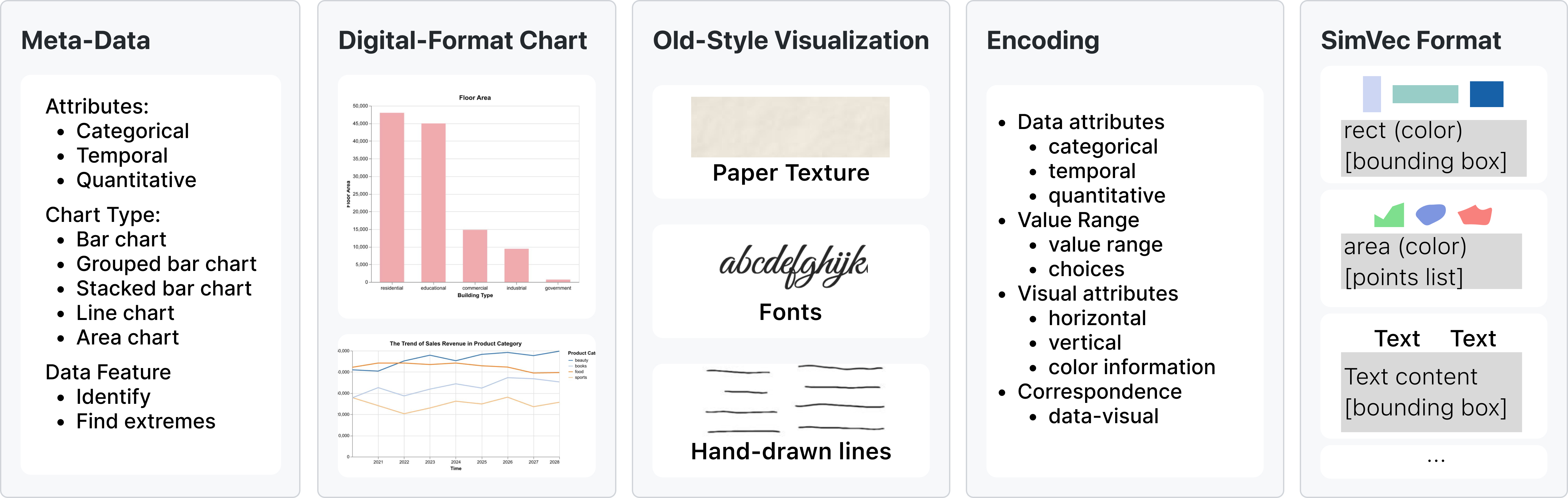}
    \caption{
    We convert the metadata into digital-format charts and then into historical-style charts. We also generate corresponding SimVec representations for the charts.}
    \label{fig:overview}
\end{figure}

\section{SimVec Format}

Vector formats (e.g., SVG) are widely used to represent visualizations.
However, their structural flexibility and stylistic richness can hinder machine understanding.
First, SVG supports nested \texttt{<g>} groups with transform attributes, leading to structural variability—semantically identical layouts may differ significantly in representation.
Second, SVG provides multiple encoding methods for the same visual element.
For example, a simple bar can be represented using a \texttt{rect}, \texttt{path}, \texttt{polygon}, or \texttt{points} element.
Although these alternatives render identically to the human eye, they introduce syntactic inconsistency that complicates automated parsing and reasoning.
Moreover, SVG files often include stylistic metadata, such as \texttt{font-family}, filters, or shadows, that are not essential for understanding the underlying data.
To address these issues, we propose \textbf{SimVec}, a simplified vector format designed to retain the essential visual structure while enforcing a consistent, machine-readable representation.
SimVec reduces complexity by:  
(1) flattening nested elements into an ordered list,  
(2) standardizing coordinates and color encodings, and  
(3) removing redundant styling.
To support a variety of visualizations (\textbf{C1}), SimVec consists of four element types, as depicted in \autoref{tab:simvec_elements}.
SimVec is compact, reducing the token count by about 90\% compared to the original SVG chart (e.g., generated using Vega-Lite).
The tokenized length of the SVG may exceed the context window limits of many MLLMs, making it difficult for models to process.

\begin{table}[ht]
    \centering
    \tiny
    \begin{tabular}{p{0.08\columnwidth}|p{0.22\columnwidth}|p{0.2\columnwidth}|p{0.28\columnwidth}}
        \hline
        \textbf{Element} & \textbf{Description} & \textbf{Format} & \textbf{Example} \\
        \hline
        Text & 
        Represents textual content with position and styling information. Used for titles, labels, and annotations & 
        \texttt{\{text: content, bbox: [left, top, width, height], color: ($h$, $s$, $l$)\}} & 
        \texttt{\{text "Title" [100, 50, 200, 30] hsl (0, 0, 18)\}} \\
        \hline
        Rectangle & 
        Used for bars, backgrounds, and other rectangular shapes. Defined by bounding box coordinates & 
        \texttt{\{rect: bbox[left, top, width, height], color: ($h$, $s$, $l$)\}} & 
        \texttt{\{rect [100, 100, 50, 150] hsl (10, 15, 12)\}} \\
        \hline
        Line & 
        Represents axes, grid lines, and connecting lines. Defined by a series of points & 
        \texttt{\{line: points[($x_1$, $y_1$), ($x_2$, $y_2$), ...], color: ($h$, $s$, $l$)\}} & 
        \texttt{\{line [(0, 0), (100, 100)] hsl (0, 0, 5)\}} \\
        \hline
        Polygon & 
        Used for complex shapes and areas. Defined by a series of connected points forming a closed shape & 
        \texttt{\{polygon: points[($x_1$, $y_1$), ($x_2$, $y_2$), ...], color: ($h$, $s$, $l$)\}} & 
        \texttt{\{polygon [(0, 0), (50, 50), (100, 0)] hsl (5, 10, 15)\}} \\
        \hline
    \end{tabular}
    \caption{
        All the coordinates and size mentioned above are described using a uniform value where the size is set to 1000.
        The color is represented in HSL color space and uniformized to [0-20] range.
    }
    \label{tab:simvec_elements}
\end{table}

\section{Dataset Construction}

To enable models to effectively understand charts, 
there are four design considerations for the \datasetname{} dataset:

    \textbf{C1: Diverse Visualization Types.}
    The dataset should include diverse visualization types and data attributes.
    Specifically, the dataset should contain common visualization types~\cite{battle2018beagle} such as bar charts, line charts, and area charts, covering different attributes like numerical, categorical, and temporal attributes from diverse topics.
    
    \textbf{C2: Accurate Data Features.}
    We focus on data-centric QA tasks, specifically retrieving values and finding extremes~\cite{amar2005tasks}.
    Unlike trend detection, correlation estimation, or outlier detection, which can often be inferred from the overall shape of the chart, data QA requires precise decoding of the underlying quantitative values.
    
    \textbf{C3: Intermediate Reasoning.}
    \revision{Many data QA questions require step-by-step reasoning—for instance, interpreting axis scales before mapping visual elements to values~\cite{pinker1990graph}.
    \datasetname{} supports this process through chain-of-thought (CoT) annotations.}

    \textbf{C4: Robustness to Imperfect Visual Inputs.}
    To enhance practical relevance, \datasetname{} incorporates charts from realistic settings, including hand-drawn visualizations sourced from historical documents. These visualizations often lack original data and feature noisy, irregular layouts, presenting unique challenges for perception and reasoning.
    \revision{Incorporating such visualizations can benefit the development of models capable of handling a wide range of visualizations beyond clean synthetic charts.}

\begin{figure}
    \centering
    \includegraphics[width=\columnwidth]{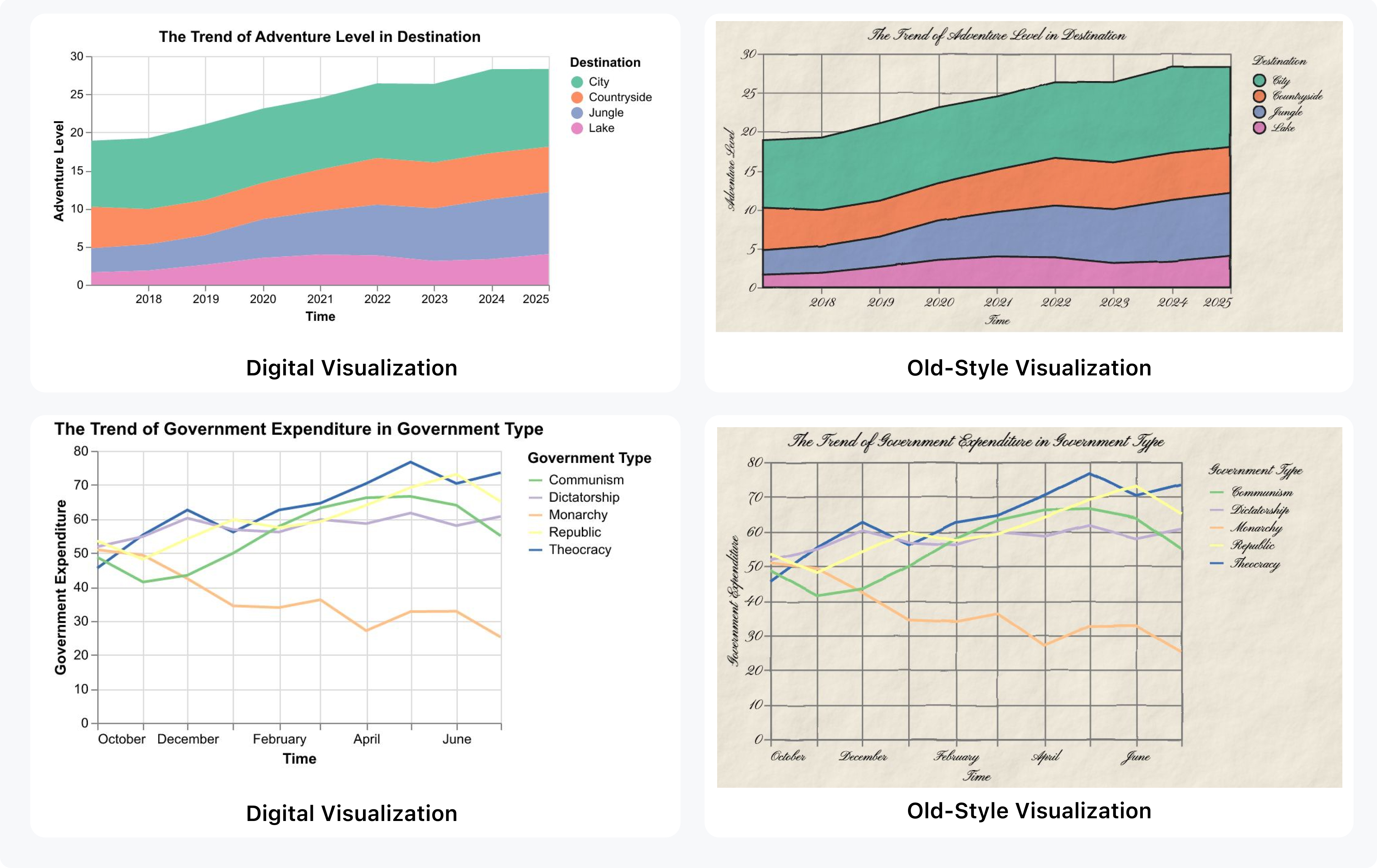}
    \caption{Historical-style visualizations with paper-texture, hand-drawn fonts, and hand-drawn lines.}
    \label{fig:oldstylevis}
\end{figure}

\subsection{\datasetname{} Dimensions}

\autoref{fig:overview} illustrates the overall \datasetname{} construction pipeline.
Each instance contains the following components: a visualization image, its corresponding SimVec, and a set of question-answer (QA) pairs.
Each QA pair is accompanied by a CoT description.

\textbf{Visualization Bitmap Image.}
To ensure diversity, we prompt GPT-4o to generate meaningful data attributes based on commonly discussed topics.
For example, in the energy domain, the LLM generates a categorical attribute (e.g., energy source), a temporal attribute (e.g., year), and corresponding quantitative values.
We then populate these attributes with randomly synthesized data.
The resulting datasets are then visualized using predefined templates for bar, line, or area charts.
Color schemes are randomly assigned.

\textbf{SimVec.}
The SVG format is then converted into a SimVec format, which captures a structured and vectorized representation of the visualization content.
Ideally, if the elements of a visualization image can be reconstructed using SimVec, it will enable a precise extraction of axes and element position, size, and color.
Therefore, our dataset includes SimVec for reconstructing the visualization structure and supporting further QA tasks (\textbf{C2}).

\textbf{QA-pairs with CoT Descriptions.}
Given the metadata associated with each chart, we generate question-answer (QA) pairs. Among the low-level tasks~\cite{amar2005tasks}, we focus on data tasks (\textbf{C2}), such as retrieving values and finding extremes. A question can be: ``What is the proportion of gas in 2020?'', with the corresponding answer being ``35\%''.
Each QA pair has a CoT description.
When humans extract precise information from visualizations, they engage in an intermediate reasoning process (e.g., identifying axis mapping).
Unlike previous datasets~\cite{wei2024mchartqa}, \revision{our dataset incorporates reasoning steps in the question-answering process} (\textbf{C3}), implementing what is known as the CoT approach~\cite{wei2022chain}.
To support step-by-step reasoning, we utilize axis metadata to guide the CoT process. For bar charts, this involves identifying the target bar and mapping its height to the corresponding value using the axis scale. A typical CoT trace includes the axis scale information, intermediate calculations, and the final answer.
An example of a CoT description is
\textit{``For the Y-axis of the chart maps from 50 pixels to 450 pixels, corresponding to a percentage range of 0\% to 100\%.
The height of the bar representing Gas in 2020 is 140 pixels.
Thus, Gas in 2020 accounts for 
$ (140/(450 - 50))\times100= 35\% $.''}

\subsection{Preliminary Attempt on Historical Visualizations}

While existing datasets focus on digitally rendered charts~\cite{wei2024mchartqa,kahou2017figureqa} that are created using toolkits (e.g., Vega-Lite), historical visualizations present distinct challenges in analysis and interpretation.
\revision{Vectorization and data extraction from historical visualizations~\cite{Zhang2021MI3} require heavy user involvement.
As a preliminary step toward addressing these challenges, we mocked historical
visualizations from digitally generated charts, aiming to expose models to stylistic variability beyond modern formats and improve their generalization and recognition abilities (\textbf{C4}).}
\autoref{fig:oldstylevis} compares digital visualizations with their corresponding historical-style representations.
We employ the following steps to mock historical-style visualizations:

\begin{itemize}
    \item \textbf{Paper Texture.} 
    We simulate historical paper by incorporating natural textures and aging effects, including subtle surface irregularities and a slight yellow tint.
    These features enhance the visual distinction from modern, digitally rendered charts.
    
    \item \textbf{Hand-drawn Fonts.} 
    We utilize specialized fonts that mimic handwriting styles seen in historical documents.
    These fonts provide a visually authentic handwritten appearance.
    
    \item \textbf{Hand-drawn Lines.} 
    We reproduce the natural imperfections of manually drawn lines by introducing controlled variations in thickness and direction. These variations result in irregularities characteristic of hand-drawn charts.
\end{itemize}

\datasetname{} comprises 2,999 visualizations, including 1,012 bar charts, 1,012 line charts, and 975 area charts. Each visualization is accompanied by a corresponding encoding output and a SimVec representation.
The dataset includes 2,999 identification tasks and 5,642 extreme value detection tasks.

%% file: sections/5_experiments.tex
\section{Experiments}

To evaluate whether \datasetname{} brings measurable improvements in visualization understanding, we conducted experiments centered on three key hypotheses, which examine: (1) the limitations of existing MLLMs, (2) the potential of CoT and SimVec to enhance visualization understanding, and (3) whether such gains are attributable to SimVec's ability to support visualization reconstruction.

\begin{itemize}
    \item \textbf{H1:} Existing MLLMs cannot accurately support data understanding tasks without fine-tuning using \datasetname{}.
    \item \textbf{H2:} \revision{CoT reasoning improves data QA accuracy, and is further enhanced with SimVec support.}\removed{Incorporating CoT and SimVec strategies enhances model accuracy on open-source models.}
    \item \textbf{H3:} Training MLLMs with SimVec representations improves their ability to reconstruct visualizations by providing a compact and structured encoding of visual elements.

\end{itemize}

\begin{table}[ht]
    \centering
    \small
    \caption{
       Accuracy is reported as the percentage of predictions with deviations of less than 5\%, 10\%, and 20\% of ground truth values.
       Qwen-VL did not benefit from SimVec, likely due to its limited ability to accurately localize chart elements, which may have introduced additional computational burden during training.
    }
    \begin{tabular}{@{}lrrr@{}}
        \toprule
        \textbf{Model}                     &  $<$ 5\%  & $<$ 10\% & $<$ 20\% \\ \midrule
        
        GPT-4o                   & 16.54\%  & 29.62\%    & 42.69\%   \\ 
        MiniCPM (zero-shot)          & 11.92\%  & 17.69\%   & 57.69\%  \\
        DeepSeek-VL(zero-shot)            & 10.00\%  & 17.31\%    & 26.92\%   \\
        Qwen-VL (zero-shot)          & 7.31\%  & 13.46\%   & 21.15\%   \\ \midrule
        MiniCPM (SimVec + QA w/ CoT)  & \underline{\textbf{53.84\%}}   & \underline{\textbf{69.23\%}}  & \underline{\textbf{80.77\%}}  \\
        MiniCPM (QA w/ CoT)           & 29.23\% & 45.76\%   & 69.23\%   \\
        MiniCPM (QA w/o CoT)          & 26.92\%  & 41.92\%   & 25.38\%  \\
        Qwen-VL (SimVec + QA w/ CoT) & 5.38\% &  10.00\%  &  18.08\%  \\
        Qwen-VL (QA w/ CoT)           & 12.31\%  & 21.54\%    & 35.77\%   \\
        Qwen-VL (QA w/o CoT)          & 11.54\%  & 19.62\%   & 31.15\%   \\
        \bottomrule
    \end{tabular}
    \label{tab:model_comparison}
\end{table}

\begin{table*}[ht]
    \centering
    \small
    \caption{
        Percentage of answers that have a difference rate under 5\%, 10\%, and 20\%.
        The best performance is achieved by MiniCPM fine-tuned with chain-of-thought supervision and SimVec.
    }
    \setlength{\aboverulesep}{0pt}
    \setlength{\belowrulesep}{0pt}
    \begin{tabular}{@{}l|rrr|rrr|rrr@{}}
        \toprule
        \multirow{2}{*}{\textbf{Model}} & \multicolumn{3}{c|}{\textbf{Area Chart}} & \multicolumn{3}{c|}{\textbf{Bar Chart}} & \multicolumn{3}{c}{\textbf{Line Chart}} \\ 
        \cline{2-10}
        & $<$ 5\% & $<$ 10\% & $<$ 20\% & $<$ 5\% & $<$ 10\% & $<$ 20\% & $<$ 5\% & $<$ 10\% & $<$ 20\% \\ 
        \midrule

        MiniCPM (SimVec + QA w/ CoT) & \underline{\textbf{38.16\%}} & \underline{\textbf{55.26\%}} & \underline{\textbf{69.74\%}} & \underline{\textbf{46.91\%}} & \underline{\textbf{65.43\%}} & \underline{\textbf{80.25\%}} & \underline{\textbf{70.87\%}} & \underline{\textbf{82.52\%}} & \underline{\textbf{89.32\%}} \\ 
        MiniCPM (QA w/ CoT) & 18.42\% & 39.47\% & 63.16\% & 33.33\% & 45.68\% & 74.07\% & 33.98\% & 50.49\% & 69.90\% \\ 
        MiniCPM (QA w/o CoT) & 23.68\% & 36.84\% & 51.85\% & 20.99\% & 34.57\% & 51.85\% & 33.01\% & 51.46\% & 64.08\% \\ 
        MiniCPM (zero-shot) & 2.63\% & 6.58\% & 14.47\% & 17.28\% & 22.22\% & 25.93\% & 14.56\% & 22.33\% & 33.01\% \\ 
        
        \bottomrule
    \end{tabular}
    \label{tab:chart_type_comparison}
\end{table*}

\subsection{Performance Comparison of Zero-Shot Models}

To validate \textbf{H1}, we evaluated the performance of various MLLMs in a zero-shot setting.
We selected the leading closed-source MLLM, GPT-4o, as well as several state-of-the-art open-source MLLMs: MiniCPM~\cite{hu2024minicpm}, DeepSeek-VL~\cite{lu2024deepseekvl}, and Qwen-VL~\cite{bai2023qwenvl}.
In our experiment, data tasks rely on accurately localizing individual items.
However, due to the lack of nominal references in scatter plots, pinpointing and directly identifying individual points is relatively challenging.
Therefore, we selected bar charts, area charts, and line charts, as they allow for clearer localization of reference points through category and label information.
These questions necessitate decoding the visual encoding in the chart to derive specific numerical values.
The evaluation results are presented in \autoref{tab:model_comparison}.
The evaluation metrics are presented as percentages, where accuracy is measured at three threshold levels: within 5\%, 10\%, and 20\% deviation from the ground truth values.
Among the four untrained models, GPT-4o performed best, which aligns with expectations for a high-capacity model.
MiniCPM outperformed Qwen-VL and DeepSeek-VL, likely due to its specialized training in text localization, which enhanced its ability to directly extract numerical values.
\removed{However, the overall performance of these models was moderate.
They primarily extracted values directly from charts but often deviated significantly from the correct values.}

\revision{\textbf{H1} is supported. The overall performance of these MLLMs was moderate: they failed on tasks that required fundamental reasoning or simple calculations.}
In the following experiments, MiniCPM is chosen as the primary model, and Qwen-VL is used as a baseline.

\subsection{Influence of CoT and SimVec on Data QA Accuracy}
\label{sec:training_datasets}

To validate \textbf{H2}, we compare three settings used for model training.
We fine-tuned the model on 8 A100 40GB GPUs in about 14 hours (Taking MiniCPM for example).
When addressing the same data question, the responses differ in the three settings:
\textbf{(1) QA without CoT}: This setting takes a visualization image and a question as input and directly provides a numerical answer.
\textbf{(2) QA with CoT}: Compare to the direct answer setting, we add CoT description prior to the numerical answer.
\textbf{\revision{(3) SimVec + QA with CoT}\removed{Enhanced with SimVec and CoT}}:
\revision{In addition to the CoT, the model is additionally trained to predict the SimVec representation of the chart.}
\removed{The model is trained to process SimVec information through a question-answering method, where the question is \textit{``What is the SimVec of this chart?''} and the answer is the SimVec of the input chart.}

\autoref{tab:model_comparison} shows that the model utilizing CoT supported by SimVec yields optimal performance.
Training enhanced with CoT significantly outperforms the direct answer setting.
For MiniCPM, the integration of SimVec information significantly enhanced its accuracy compared to using CoT reasoning alone.
\revision{SimVec encapsulates visual attributes of text and marks, which provides the information required for CoT (i.e., the axes and visual attributes).}
\removed{The SimVec format encapsulates both textual and spatial information, enabling precise mapping of $X$ and $Y$ axes components and visual attributes of chart elements.}
However, for Qwen-VL, SimVec implementation did not yield improved accuracy, due to the model's limited capability in chart element localization, and SimVec information may have introduced unnecessary computational overhead for Qwen-VL.
A detailed analysis of performance across various chart types for MiniCPM is illustrated in \autoref{tab:chart_type_comparison}.
Notably, MiniCPM (SimVec + CoT) surpasses all other model configurations in all chart types.
The model performs substantially better on line charts compared to bar and area charts. This may be because line charts convey values more directly through position, whereas bar and area charts rely on height, which may involve stacking and thus introduce additional visual complexity.
\revision{Despite improvements from SimVec and CoT, challenges remain.
Current MLLMs often struggle to estimate spatial properties like bar height or line position, especially without explicit labels.
Errors in early reasoning steps, such as axis decoding, tend to propagate and affect final answers.}

\textbf{H2} is supported. CoT reasoning, especially when combined with SimVec, notably enhances MiniCPM’s accuracy and achieves superior performance over current state-of-the-art MLLMs.

\begin{table}[ht]
    \centering
    \small
    \caption{Reconstruction Quality for Different Chart Types. The distance unit is 1/1000 of the image size.
    }
    \begin{tabular}{lrrr}
        \hline
        \textbf{Index} & \textbf{Line} & \multicolumn{1}{r}{\textbf{Bar}} & \multicolumn{1}{r}{\textbf{Area}} \\ \hline
        Text Hit Rate             & 99.79\% & 99.60\% & 99.83\% \\
        Text Similarity           & 98.37\% & 96.60\% & 98.72\% \\
        Text Center Distance      & 2.89    & 8.26    & 2.70    \\
        Element Color Distance    & 1.06    & 1.78    & 2.14    \\
        Element Position Distance & 8.76    & 10.11   & 29.26   \\ \hline
        \end{tabular}
    \label{tab:recover_distance_comparison}
\end{table}

\subsection{Reconstruction Capability using SimVec Format}

\begin{figure}[ht]
    \centering
    \includegraphics[width=\columnwidth]{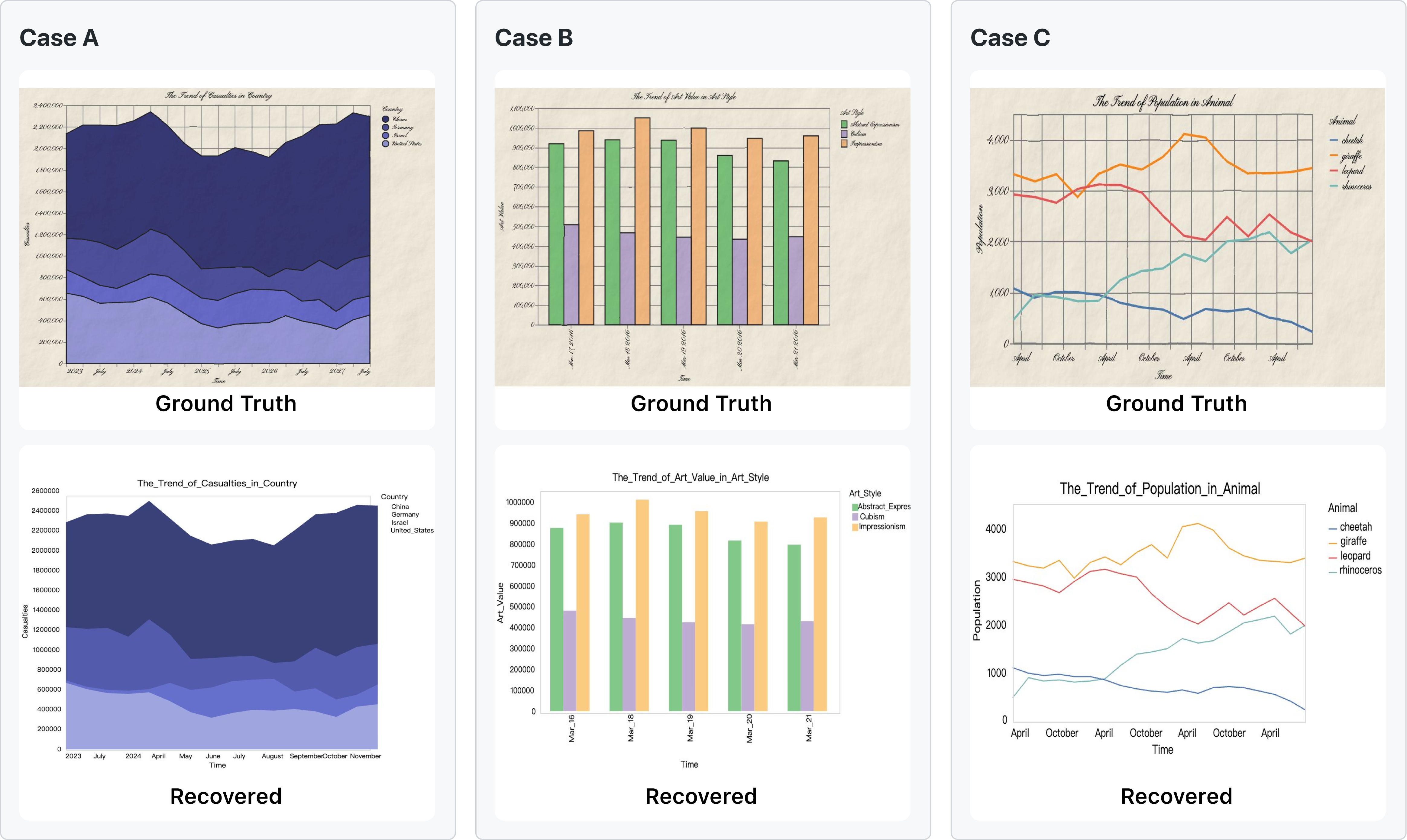}
    \caption{For each case (A, B, and C), the top panel displays the original input visualization, while the bottom panel shows the reconstructed result rendered using the SimVec output by the model.
    }
    \label{fig:recover_case}
\end{figure}

To evaluate hypothesis \textbf{H3}, we assess the reconstruction capabilities using the SimVec format.
We use MiniCPM (SimVec+CoT) to take an image as input and generate the corresponding SimVec as output.
The dataset includes 100 images of bar, line, and area charts.
\autoref{fig:recover_case} shows the original input image and the image rendered using the output reconstructed SimVec.
The results demonstrate that different types of charts can be recovered to a satisfactory extent.
As shown in \autoref{tab:recover_distance_comparison}, 
we calculate the hit rate and similarity as percentages. 
The distance is calculated in terms of the average number of pixels, where the image size is normalized to 1000 pixels.
The quantitative tests include the text accuracy and graphics accuracy:

\textbf{Text Accuracy Evaluation.}
To assess the model's text reconstruction capabilities, we employed multiple metrics: the \textit{text hit rate} to measure the proportion of successfully recovered text elements, the \textit{text similarity} using Levenshtein distance~\cite{yujian2007normalized}, and the \textit{text center distance} to evaluate spatial accuracy of text placement.
Experimental results demonstrate the model's robust performance, with text similarity reaching 98\%.
The average center distance deviation ranges from 0.27\% to 0.83\% of the image size (defined as the larger dimension of height or width), indicating high precision in spatial text recovery.

\textbf{Graphics Accuracy Evaluation.}
We employed the average pixel distance between predicted vertices and their corresponding ground truth as a metric to assess the reconstruction capabilities.
The line elements in line charts demonstrated the highest precision with an average positional deviation of merely 0.88\% of the image size.
Bar elements showed comparable accuracy at 1\% of image size, while area charts exhibited a slightly higher average distance of approximately 3\%.
These performance metrics align consistently with the data accuracy rankings presented in \autoref{tab:chart_type_comparison}.
For color fidelity assessment, we calculated the Euclidean distance between predicted and ground truth colors in the HSL color space (each dimension normalized to [0, 20]).
The color differences' effect on overall perception is minor; for example, the differences are sufficient to identify distinct colors but unlikely to impair understanding.

These results validate \textbf{H3}, confirming that the SimVec format effectively supports high-fidelity reconstruction of both textual and graphical elements.
Nevertheless, errors at the value level persist, mainly due to error accumulation in multi-step reasoning.

%% file: sections/6_conclusion.tex
\section{Conclusion and Future Work}

\revision{
We introduce \datasetname{} that pairs bitmap charts with their SimVec representations and includes data QA tasks with CoT description, enabling supervised training for visualization understanding.
We aim to expand the dataset to include infographics to support broader visualization scenarios.}